\documentclass[fleqn,12pt,twoside]{article}
\usepackage{espcrc1}
\usepackage{natbib}
\bibpunct{[}{]}{,}{a}{}{,}

\usepackage{graphicx}
\usepackage[figuresright]{rotating}

\newcommand{\AmS}{{\protect\the\textfont2
  A\kern-.1667em\lower.5ex\hbox{M}\kern-.125emS}}
\newcommand{\la}{\mathrel{\vcenter
     {\offinterlineskip \hbox{$<$}\hbox{$\sim$}}}}

\makeatletter
\renewcommand{\@biblabel}[1]{#1.}
\makeatother

\title{The mechanism of core-collapse supernovae and the 
       ejection of heavy elements}

\author{H.-Th. Janka\address[MPA]{Max-Planck-Institut f\"ur Astrophysik,\\
        Karl-Schwarzschild-Str.~1, D-85741 Garching, Germany},
        R. Buras\addressmark 
        ~and
        M. Rampp\addressmark}
       
\begin{document}

\maketitle

\begin{abstract}
We present here the first results of two-dimensional hydrodynamical
simulations of the neutrino-heating phase in the collapsed core of 
a 15$\,M_{\odot}$ star, where the
neutrino transport is treated with a variable Eddington factor method
for solving the Boltzmann transport equation, and the neutrino
interactions include nucleon-nucleon bremsstrahlung, nucleon recoils
and correlations, and weak-magnetism effects as well as direct 
interactions between neutrinos of different flavors. With the given 
input physics (neutrino reactions and nuclear equation of state),
our best simulations do not develop strong convection in the 
neutrino-heating layer behind the shock
and do not yield explosions. With about 30\% higher neutrino-energy
deposition behind the shock, however, an explosion occurs on a timescale
of 150$\,$ms after core bounce. It leaves behind a neutron star with
an initial baryonic mass of 1.4$\,M_{\odot}$ and ejects the $N=50$
isotopes of Sr, Y and Zr in amounts consistent with Galactic abundances.
\end{abstract}

\section{INTRODUCTION}
\label{sec:intro}

Wilson and collaborators (\cite[]{wilmay88}--\cite[]{totsat98})
have routinely obtained delayed supernova explosions by the 
neutrino-heating mechanism~\cite[]{betwil85}
for more than ten years in one-dimensional simulations.
In order to get a sufficiently large neutrino-energy 
deposition for reviving the stalled shock, they 
{\em assumed} that neutron-finger convective instabilities 
boost the neutrino luminosities from the hot neutron star. 
Moreover, Mayle et.~al. \cite{maytav93} used a special nuclear
equation of state with
a high abundance of pions in the nuclear matter to reproduce
the explosion energy of SN~1987A. Both assumptions are
not generally accepted. Therefore investigations of the
neutrino-driven mechanism should be continued with making
less controversial postulats about the involved physics.

Forced by observations of large-scale anisotropies and 
mixing in SN~1987A and directed by the fact that the 
neutrino-heating layer is convectively unstable \cite[]{bet90},
supernova modelers began to perform multi-dimensional
simulations of the post-bounce supernova evolution
and discovered that violent motions set in on
a timescale relevant for the delayed mechanism 
(\cite[]{herben92}--\cite[]{shiebi01}).
In fact, increasing the postshock pressure by rising,
outward pushing neutrino-heated matter and allowing for
simultaneous accretion (thus maintaining
high accretion luminosities) and shock expansion (thus
enlarging the layer of neutrino-energy deposition), this
postshock convection turned out to have a very helpful 
influence on the explosion: Models which failed to blow up in
spherical symmetry were found to succeed in two dimensions.

Making use of this effect and employing
an advanced version of the Herant et.~al. code \cite{herben94},
Fryer has started to explore a variety of progenitors 
to determine whether neutron stars or black holes are left behind
as compact remnants \cite{fry99}. Fryer and Heger expanded the study
to rotating stars \cite{fryheg00}, and most recently Fryer and Warren 
have finished the first full supernova simulations in 
three dimensions, finding good agreement with results that were
obtained in the two-dimensional case \cite{frywar02}.

Though these simulations are undoubtedly pioneering first 
steps, they suffer from a major
uncertainty: The transport of neutrinos in the supernova
core is described in a much simplified way, using grey
flux-limited diffusion and ignoring Doppler-shift and 
gravitational redshift effects on the neutrino spectrum.
This approximation is too crude to allow for serious
conclusions on the explosion mechanism. The importance of
an accurate numerical description of the neutrino transport
has repeatedly been pointed out by Bruenn and by Mezzacappa and
collaborators (e.g., \cite[]{mesmez98}--\cite[]{liemes02}).

For this reason, a new generation of supernova codes has
been developed in the past years with the aim to replace
the widely used (multi-group) flux-limited 
diffusion treatments (e.g., \cite[]{may85}--\cite[]{coobar92})
by a direct solution of the Boltzmann transport equation
(\cite[]{yamjan99}--\cite[]{ramjan00},\cite[]{mesmez98,mezlie01}).
Progress was also made for
including the effects of general relativity (GR) in a
rigorous way in neutrino-hydrodynamics codes in spherical
symmetry (\cite[]{liemez01}--\cite[]{liemes02:code}). Although 
improved flux limiters for the neutrino transport \cite[]{jan92}
or two-moment closure schemes \cite[][ and references
therein]{smicer97} have been suggested and may work better than
previously used flux limiting prescriptions (for an 
apparently successful new development in this direction, 
see the results by Bruenn shown in Ref.~\cite{liemes02:code}), 
any such approximation has ultimately to be 
compared with Boltzmann solvers applied in dynamical 
supernova calculations. 

In this conference contribution we shall summarize our efforts
to combine a Boltzmann solver for the neutrino transport with
a multi-dimensional hydrodynamics code, using a state-of-the-art
description of neutrino-matter interactions and an
approximation of GR which is (hopefully)
sufficiently accurate to model supernova explosions
and neutron star formation. First
results of two-dimensional simulations with this new code
named MuDBaTH ({\bf Mu}lti-{\bf D}imensional {\bf B}oltzm{\bf a}nn
{\bf T}ransport and {\bf H}ydrodynamics) will also be presented.

\section{A NEW RADIATION-HYDRODYNAMICS CODE FOR NEUTRINOS}
\label{sec:code}

For the integration of the equations of hydrodynamics we employ the
Newtonian finite-volume code PROMETHEUS \cite[]{frymue89}, which was 
supplemented by additional
problem specific features by Keil \cite{kei97}.  
PROMETHEUS is a direct Eulerian, time-explicit implementation of the
Piecewise Parabolic Method (PPM) of Colella and Woodward \cite{colwoo84}.

The neutrino transport is done with the Boltzmann solver scheme
that is described in much detail in Ref.~\cite{ramjan02}. The
integro-differential character of the Boltzmann equation is
tamed by applying a variable Eddington factor closure to the 
neutrino energy and momentum equations (and the simultaneously 
integrated first and second order moment equations for neutrino
number). The variable Eddington factor
is determined from the solution of the Boltzmann equation and
the system of Boltzmann equation and its moment equations is
iterated until convergence is achieved. Employing this scheme in
multi-dimensional simulations in spherical coordinates, we solve the 
moment equations on the different angular bins of the numerical grid but 
calculate the variable Eddington factor only once on an angularly
averaged stellar background. This approximation is good only for
situations without significant global deformations. Since
the iteration of the Boltzmann equation has to be done only 
once per time step, appreciable
amounts of computer time can be saved \cite[]{ramjan02}.
We point out here that it turned out to be 
necessary to go an important step beyond this simple ``ray-by-ray''
approach. Physical constraints, namely the conservation of lepton
number and entropy within adiabatically moving fluid elements,
and numerical requirements, i.e., the stability of regions
which should not develop convection according to a mechanical stability
analysis, make it necessary to take into account the coupling of 
neighbouring rays at least by lateral advection terms and neutrino 
pressure gradients.

General relativistic effects are treated only approximately in
our code \cite{ramjan02}. The current version contains a modification 
of the gravitational potential by including correction terms due to
pressure and energy of the stellar medium and neutrinos, which
are deduced from a comparison of the Newtonian and relativistic
equations of motion. The neutrino transport contains gravitational
redshift and time dilation, but ignores the distinction between
coordinate radius and proper radius. This simplification is 
necessary for coupling
the transport code to our basically Newtonian hydrodynamics. 
Although a fully relativistic treatment would be preferable, tests
showed that these approximations seem to work reasonably well as
long as there are only moderate deviations ($\sim$10--20\%) of 
the metric coefficients from unity and the infall velocities do
not reach more than 10--20\% of the speed of light.

As for the neutrino-matter interactions, we discriminate between
two different sets of input physics. On the one hand we have 
calculated models with conventional (``standard'') neutrino 
opacities, i.e.,
a description of the neutrino interactions which follows closely 
the one used by Bruenn and Mezzacappa and collaborators
\cite[]{bru85,mezbru93:coll,mezbru93:nes}. It assumes
nucleons to be uncorrelated, infinitely massive scattering
targets for neutrinos. In these reference runs we have 
usually also added
neutrino pair creation and annihilation by nucleon-nucleon
bremsstrahlung \cite[]{hanraf98}. Details of our
implementation of these neutrino processes can be found in 
Ref.~\cite[]{ramjan02}. 

A second set of models was computed with
an improved description of neutrino-matter interactions. 
Besides including nucleon thermal motions and recoil, which
means a detailed treatment of the reaction kinematics 
and allows for an accurate evaluation of nucleon phase-space
blocking effects, we take into account nucleon-nucleon
correlations \cite[following Refs.][]{bursaw98,bursaw99},
the reduction of the nucleon effective mass, and the
possible quenching of the axial-vector coupling
in nuclear matter \cite[]{carpra02}. In addition, we have 
implemented weak-magnetism corrections as described in 
Ref.~\cite{hor02}. The sample of neutrino processes was
enlarged by also including scatterings of muon and tau
neutrinos and antineutrinos off electron neutrinos and
antineutrinos and pair annihilation reactions between
neutrinos of different flavors 
(i.e., $\nu_{\mu,\tau} + \bar\nu_{\mu,\tau}
\longleftrightarrow \nu_e + \bar\nu_e$; \cite{burjan02:nunu}).

Our current supernova models are calculated with the nuclear 
equation of state of Lattimer and Swesty \cite{latswe91}. For the
density regime below $6\times 10^7\,$g/cm$^3$
we switch to an equation of state that considers
electrons, positrons and photons, and nucleons and nuclei with
an approximative treatment for composition changes due to nuclear
burning and shifts of nuclear statistical equilibrium \cite[]{ramjan02}.

\section{RESULTS: NEW 1D AND 2D SUPERNOVA SIMULATIONS}
\label{sec:results}

We have performed a number of core-collapse simulations
in spherical symmetry and then followed the post-bounce
evolution in one and two dimensions. All described
calculations were started from a 15$\,M_{\odot}$ progenitor 
star, Model s15s7b2, provided to us by S.~Woosley. 
Adopting the naming, we label our models by s15N for
Newtonian runs and s15G for runs with approximate treatment
of general relativity, followed by letters ``so'' when
``standard neutrino opacities'' were used and by ``io'' 
in case of our state-of-the-art improvement of the 
description of neutrino-matter interactions. The model names
have a suffix that discriminates between 1D (``\_1d'') and
2D simulations (``\_2d''). 

We have varied yet another aspect in our simulations. 
Some of the models were computed with a version of the transport
code where the velocity dependent (Doppler shift and aberration) 
terms in the neutrino momentum equation (and the corresponding
terms in the Boltzmann equation for the antisymmetric average
of the specific intensity; see Ref.~\cite{ramjan02}) were
omitted. These terms are formally of order $v/c$ and should
be small for low velocities. This simplification of the 
neutrino transport, our so-called ``Case A'',
was used in the Newtonian models and in
those models with approximate GR which have names ending with 
the letter ``a''. Models with neutrino transport including
all velocity dependent terms also in the neutrino momentum 
equation (``Case B'') can be identified by the ending ``b'' 
of their names. Table~\ref{table:1} provides an overview of
the computed models.

\begin{table}[htb!]
\caption{Input physics for our set of computed models. See text for details.}
\label{table:1}
\newcommand{\m}{\hphantom{$-$}}
\newcommand{\cc}[1]{\multicolumn{1}{c}{#1}}
\renewcommand{\tabcolsep}{1pc} 
\renewcommand{\arraystretch}{1.1} 
\begin{tabular}{@{}llllll}
\hline
Model & Dim. & Gravity & $\nu$ Reactions & Transport & Wedge$^\dagger$\\
\hline
s15Nso\_1d          & 1D & Newtonian  & standard       & Case A  &                  \\  
s15Nso\_2d          & 2D & Newtonian  & standard       & Case A  & $\pm 27^{\rm o}$ \\
s15Gso\_1d.b        & 1D & approx. GR & standard       & Case B  &                  \\
s15Gso\_1d.b$^\ast$ & 1D & approx. GR & standard$^\ddagger$ & Case B  &              \\
s15Gio\_1d.a        & 1D & approx. GR & improved       & Case A  &                  \\
s15Gio\_2d.a        & 2D & approx. GR & improved       & Case A  & $\pm 43.2^{\rm o}$ \\
s15Gio\_1d.b        & 1D & approx. GR & improved       & Case B  &                  \\
s15Gio\_2d.b        & 2D & approx. GR & improved       & Case B  & $\pm 43.2^{\rm o}$ \\
\hline
\end{tabular}\\[2pt]
$^\dagger$ Angular wedge of the spherical coordinate grid around the equatorial plane.\\
$^\ddagger$ Calculation without neutrino-pair creation by nucleon-nucleon
bremsstrahlung.
\end{table}

\begin{figure}[htb!]
\begin{minipage}[t]{0.55\textwidth}
\includegraphics[width=\textwidth]{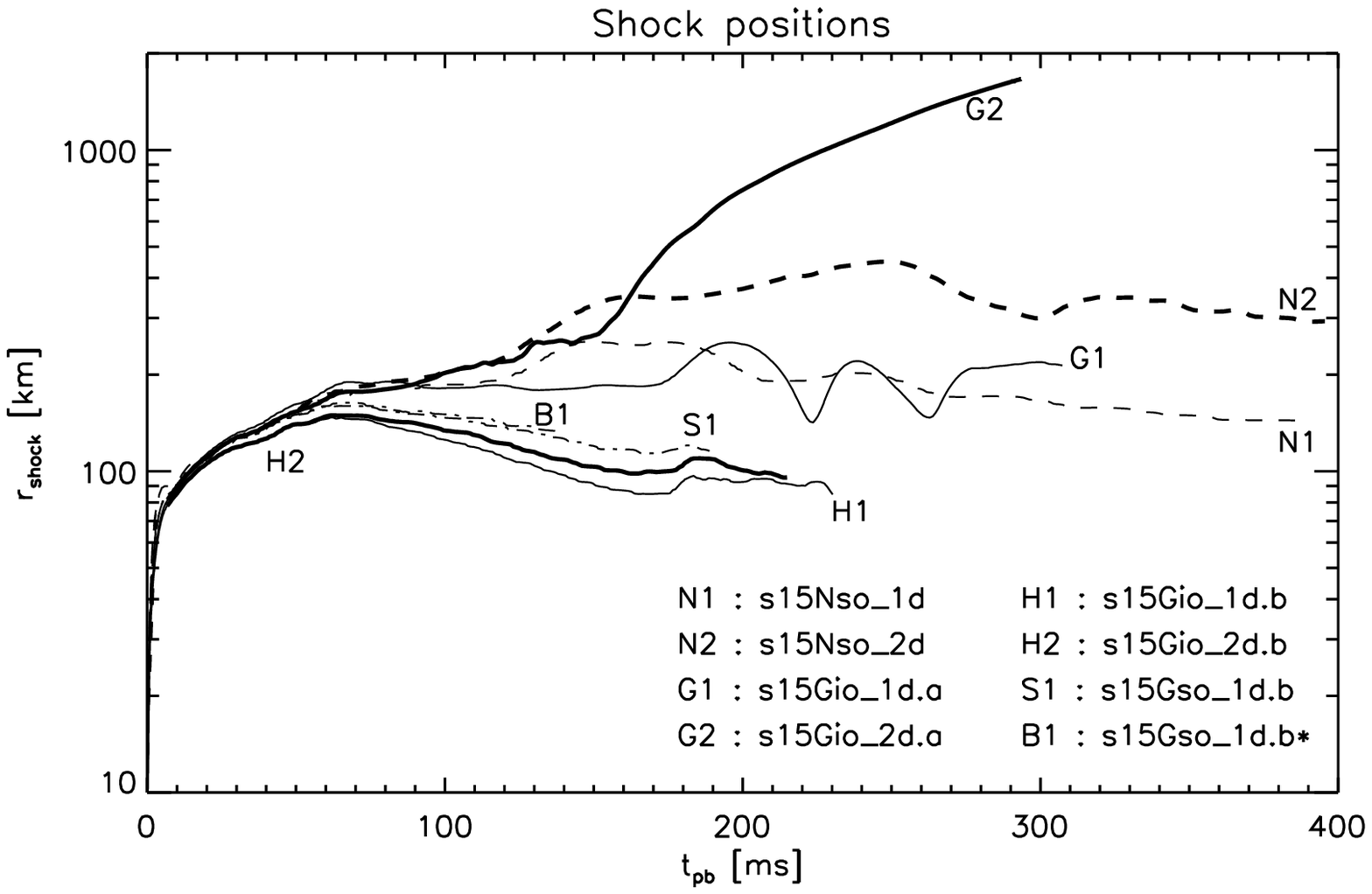}
\vspace{-15mm}
\caption{Shock trajectories for all models. Bold lines correspond
to two-dimensional simulations.}
\label{fig:shockradii}
\end{minipage}
\hspace{\fill}
\begin{minipage}[t]{0.4\textwidth}
\includegraphics[width=\textwidth]{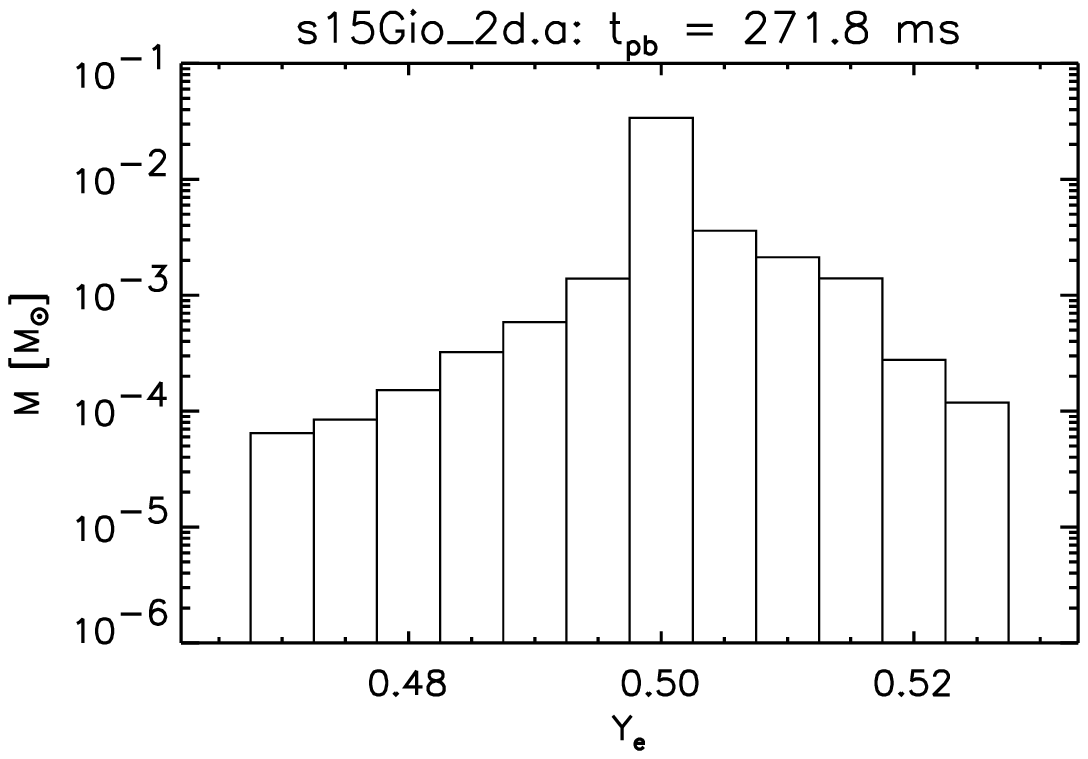}
\vspace{-15mm}
\caption{Ejecta mass vs.~$Y_e$ for Model s15Gio\_2d.a at 272$\,$ms
         post bounce (total mass: $0.041\,M_{\odot}$).}
\label{fig:yemass}
\end{minipage}
\end{figure}

The shock trajectories of all models are displayed in 
Fig.~\ref{fig:shockradii}. No explosions were obtained with the
most complete implementation of the transport equations, Case~B
(see lines labeled with H1, H2, S1, B1, corresponding to 
Models~s15Gio\_1d.b, s15Gio\_2d.b, s15Gso\_1d.b and 
s15Gso\_1d.b$^\ast$, respectively). The shock trajectories of
this sample of models form a cluster that is clearly separated 
from the models computed with the transport version of Case~A, which
generally show a larger shock radius and therefore more optimistic
conditions for explosions. Both classes of calculations differ
mainly in the (comoving frame) neutrino energy densities around 
and outside of
the neutrinosphere. In Case~A these energy densities are larger,
corresponding to somewhat smaller neutrino losses from the
cooling layer and somewhat higher neutrino energy deposition
in the heating layer behind the supernova shock. Although
the differences are moderate (typically 10--30\%, depending
on the quantity, radial position and time) during the
first 80$\,$ms after bounce, the accumulating effect 
clearly damps the shock expansion and leads to a dramatic
shock recession after this initial phase in all models of Case~B.
This demonstrates the sensitivity of the long-time evolution
of the collapsing stellar core to smaller ``details'' of the
neutrino transport.

\begin{figure}[htb!]
\includegraphics[width=150mm,clip=]{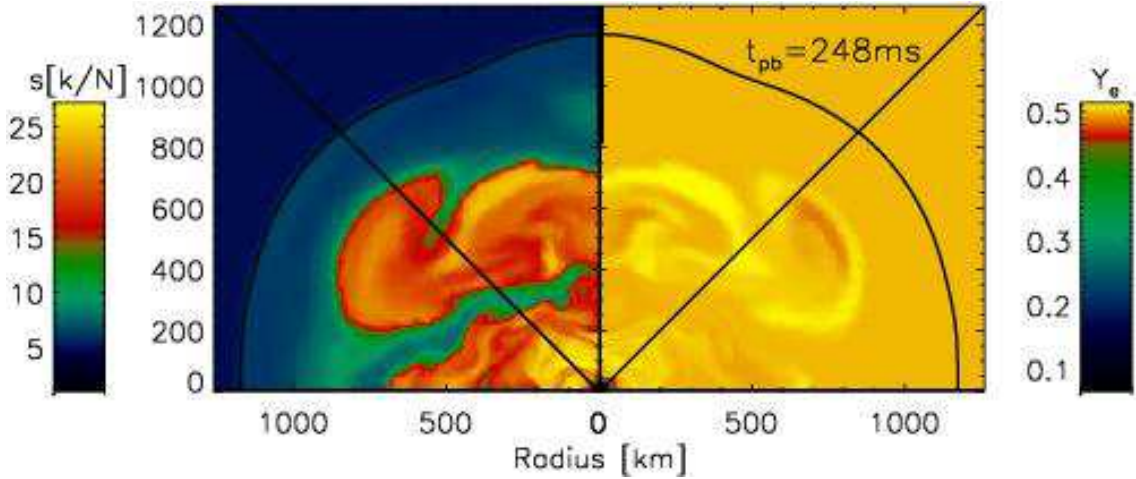}
\vspace{-8mm}
\caption{Entropy per nucleon, $s$ (left), and proton-to-nucleon
ratio, $Y_e$ (right), for exploding Model~s15Gio\_2d.a at
248$\,$ms after core bounce. The diagonal line indicates the
equatorial plane of the computational grid.}
\label{fig:snapshot}
\end{figure}

A sufficiently large shock radius for a sufficiently long
time is also crucial for the growth of convective instabilities
in the neutrino-heating layer. This is visible from 
the two-dimensional Models~s15Gio\_2d.b and
s15Gio\_2d.a. The former model fails to explode because
the convective activity behind the shock is suppressed when
the shock retreats, despite of a high energy transfer rate
to the postshock matter by neutrinos. In contrast, convective
overturn behind the shock becomes strong and is crucial for getting 
an explosion
in Model~s15Gio\_2d.a. This is obvious from a comparison with
the corresponding one-dimensional simulation, Model~s15Gio\_1d.a.
It is interesting that the Newtonian 2D run, Model~s15Nso\_2d
(with standard opacities) does marginally not explode, whereas 
Model~s15Gio\_2d.a with relativistic corrections (and 
state-of-the-art neutrino reactions) succeeds\footnote{At the
time of the conference, we had just these two 2D runs and announced
the possible success of Model~s15Gio\_2d.a. A later 2D computation with
the full neutrino moment equations (Model~s15Gio\_2d.b for Case~B),
however, then turned out to produce a dud.}. The influence
of the neutrino opacities can be directly seen by 
comparing Models~s15Gso\_1d.b and s15Gio\_1d.b. It is not
dramatic but sufficient to justify the inclusion of all the
improvements described in Sect.~\ref{sec:code}. 
Neutrino-pair creation by bremsstrahlung makes a minor
difference during the considered phases of evolution 
(Model~s15Gso\_1d.b$^\ast$ vs.\ Model~s15Gso\_1d.b).

The successfully exploding 2D model, Model~s15Gio\_2d.a, has
very interesting properties. 
Figure~\ref{fig:snapshot}
shows a snapshot at 248$\,$ms post bounce, giving the distribution
of entropy per baryon ($s$ in units of Boltzmann's constant) and
electron fraction.
The shock is located at a radius of more than
1700$\,$km at 300$\,$ms after bounce and is expanding with
about 10000$\,$km/s.
The explosion energy of this model might become rather low. It is
only $\sim 4\times 10^{50}\,$erg at that time, but
still increasing.
The explosion starts so late (at $\sim 150\,$ms post bounce)
that the proto-neutron star has accreted enough matter to have
attained an initial baryonic mass of 1.4$\,M_{\odot}$. Therefore
our simulation 
does not exhibit the problem of other successful models
which produced rather small ($\sim 1.1\,M_{\odot}$) neutron stars.
Also another problem of published explosion models 
\cite[e.g.,][]{herben94,burhay95,janmue96,fry99}
has disappeared: The ejecta mass with $Y_e \la 0.47$
is less than $10^{-4}\,M_{\odot}$ (Fig.~\ref{fig:yemass}),
thus fulfilling a constraint pointed out by 
Hoffman et.~al.~\cite{hofwoo96}
for supernovae if they should not overproduce the $N=50$ (closed neutron 
shell) nuclei, in particular $^{88}$Sr, $^{89}$Y and $^{90}$Zr,
relative to the Galactic abundances. Of course, we will have to 
follow the explosion for a longer time to make final statements
about explosion energy and ejecta composition,
and also the neutron star mass may grow by later fallback,
especially when the explosion energy remains low.

\section{SUMMARY AND CONCLUSIONS}
\label{sec:summary}

We have presented a set of core collapse and supernova
models, which were computed in spherical symmetry and in two
dimensions, using a new Boltzmann solver for the neutrino
transport and improved neutrino opacities compared to the
standard treatment in current supernova codes. We did not
find explosions, neither without nor with postshock 
convection, in our most complete simulations. 

Omitting the
velocity-dependent terms from the neutrino momentum equation,
however, leads to sufficiently large changes in the neutrino 
quantities and neutrino energy loss or deposition 
(10--30\% between neutrinosphere and shock) that very 
strong convective overturn in the neutrino-heating region
develops and drives a successful explosion. 
This demonstrates that the models are rather close
to the threshold for an explosion, causing an enormous 
sensitivity of the post-bounce evolution to an accurate 
treatment of the neutrino transport and hydrodynamics.
It also confirms the importance of postshock convection, 
because the 1D model corresponding to our exploding 2D case
does not explode. When the neutrino heating is too weak,
and the shock recedes to a small radius quickly after
its has reached its maximum radius, however, convective
activity behind the shock is suppressed and has no crucial
influence on the evolution.

We also found that Ledoux convection below the
neutrinosphere, although setting in shortly after bounce
and persistent until the end of our simulations, occurs
so deep inside the neutron star that its effects on the
$\nu_e$ and $\bar\nu_e$ luminosities and on the supernova
dynamics are insignificant.

\bigskip\noindent
{\small
{\bf Acknowledgements:} We are grateful to K.~Kifonidis, T.~Plewa
and E.~M\"uller for updates of the PROMETHEUS hydrodynamics
code, to K.~Kifonidis for contributing a matrix solver suitable for 
parallel computer platforms, to K.~Takahashi for providing routines
to calculate the improved neutrino-nucleon interactions, and to
C.~Horowitz for correction formulae for the weak magnetism.
We also thank M.~Liebend\"orfer for making output data from his 
simulations available to us for comparisons.
The described project was supported by the Sonderforschungsbereich
375 on ``Astroparticle Physics'' of the Deutsche Forschungsgemeinschaft.
The 2D simulations were only possible because a node of the
new IBM ``Regatta'' supercomputer was dedicated to this project
by the Rechenzentrum Garching. Computations were also
performed on the NEC SX-5/3C
of the Rechenzentrum Garching, and on the CRAY T90 and CRAY
SV1ex of the John von Neumann Institute for Computing (NIC) in J\"ulich.
}

\def\ref@jnl#1{{\rm#1}}
\def\apj{\ref@jnl{ApJ}}                 
\def\apjl{\ref@jnl{ApJ}}                
\def\apjs{\ref@jnl{ApJS}}               
\def\aap{\ref@jnl{A\&A}}                
\def\aapr{\ref@jnl{A\&A~Rev.}}          
\def\aaps{\ref@jnl{A\&AS}}              
\def\physrep{\ref@jnl{Phys.~Rep.}}   
\def\pra{\ref@jnl{Phys.~Rev.~A}}        
\def\prb{\ref@jnl{Phys.~Rev.~B}}        
\def\prc{\ref@jnl{Phys.~Rev.~C}}        
\def\prd{\ref@jnl{Phys.~Rev.~D}}        
\def\pre{\ref@jnl{Phys.~Rev.~E}}        
\def\prl{\ref@jnl{Phys.~Rev.~Lett.}}    
\def\jcomp{\ref@jnl{J.~Comp.~Phys.}}    

\end{document}